\documentclass[10pt,conference]{IEEEtran}
\usepackage{cite}
\usepackage{xspace} 
\usepackage{amsmath,amssymb,amsfonts}
\usepackage{algorithmic}
\usepackage{graphicx}
\usepackage{textcomp}
\usepackage{xcolor}
\usepackage{stfloats}
\usepackage{balance}
\usepackage{hyperref}
\usepackage{soul}
\usepackage{enumitem}
\usepackage{booktabs}
\usepackage{caption}
\usepackage{siunitx}
\usepackage{tabularx}
\usepackage{subcaption}
\usepackage{tcolorbox}
\usepackage{array}
\usepackage[retain-zero-uncertainty=true]{siunitx}
\tcbuselibrary{skins}
\usepackage{enumitem}
\usepackage{pifont}

\newboolean{showcomments}
\setboolean{showcomments}{true}
\ifthenelse{\boolean{showcomments}}
{\newcommand{\nb}[2] {
  \fcolorbox{black}{gray!20}{\bfseries\sffamily\scriptsize#1:}
  {\sf\small$\blacktriangleright$\textit{#2}$\blacktriangleleft$}
}
}
{\newcommand{\nb}[2]{}
}

\def\BibTeX{{\rm B\kern-.05em{\sc i\kern-.025em b}\kern-.08em
    T\kern-.1667em\lower.7ex\hbox{E}\kern-.125emX}}

\newcommand{\head}[1]{\noindent\textbf{#1.}}
\usepackage[normalem]{ulem} 

\newcommand{\changed}[1]{\textcolor{black}{#1}}

\begin{document}

\title{
A Multi-Modality Evaluation of the Reality Gap in Autonomous Driving Systems
}

\author{\IEEEauthorblockN{Stefano Carlo Lambertenghi}
\IEEEauthorblockA{
\textit{Technical University of Munich, fortiss}\\
Munich, Germany \\
stefanocarlo.lambertenghi@tum.de}
\and
\IEEEauthorblockN{Mirena Flores Valdez}
\IEEEauthorblockA{
\textit{Technical University of Munich}\\
Munich, Germany \\
mirena.flores@tum.de}
\and
\IEEEauthorblockN{Andrea Stocco}
\IEEEauthorblockA{
\textit{Technical University of Munich, fortiss}\\
Munich, Germany \\
andrea.stocco@tum.de}
}

\maketitle

\begin{abstract}
Simulation-based testing is a cornerstone of Autonomous Driving System (ADS) development, offering safe and scalable evaluation across diverse driving scenarios. However, discrepancies between simulated and real-world behavior, known as the reality gap, challenge the transferability of test results to deployed systems. In this paper, we present a comprehensive empirical study comparing four representative testing modalities: Software-in-the-Loop (SiL), Vehicle-in-the-Loop (ViL), Mixed-Reality (MR), and full real-world testing. Using a small-scale physical vehicle equipped with real sensors (camera and LiDAR) and its digital twin, we implement each setup and evaluate two ADS architectures (modular and end-to-end) across diverse indoor driving scenarios involving real obstacles, road topologies, and indoor environments. We systematically assess the impact of each testing modality along three dimensions of the reality gap: actuation, perception, and behavioral fidelity. Our results show that while SiL and ViL setups simplify critical aspects of real-world dynamics and sensing, MR testing improves perceptual realism without compromising safety or control. Importantly, we identify the conditions under which failures do not transfer across testing modalities and isolate the underlying dimensions of the gap responsible for these discrepancies. Our findings offer actionable insights into the respective strengths and limitations of each modality and outline a path toward more robust and transferable validation of autonomous driving systems.
\end{abstract}

\begin{IEEEkeywords}
autonomous driving; reality gap; virtual testing; real-world testing; vehicle-in-the-loop; mixed-reality.
\end{IEEEkeywords}

\section{Introduction}\label{sec:introduction}

To ensure the safety and reliability of autonomous driving systems (ADS) before deployment in public environments, rigorous system-level testing is indispensable~\cite{survey-lei-ma,fse-survey-robotics,icst-survey-robotics}. A common industrial practice for ADS validation follows a two-phase testing pipeline. First, ADS components-such as perception and planning modules-are trained using real-world driving data and evaluated within virtual environments via simulation-based testing, also known as simulation-in-the-loop (SiL). Subsequently, the ADS undergoes real-world testing on real vehicles on closed tracks up to public roads~\cite{Cerf:2018:CSC:3181977.3177753,10-million-miles,waymo-driver,waymos-secret-testing}.
Real-world testing, while more faithful, is costly, time-consuming, and constrained in scope and safety.
Despite their scalability, simulations cannot fully replicate real-world physical phenomena, such as sensor noise, actuator delays, and environmental complexity. The resulting mismatch is known as the \textit{reality gap}~\cite{survey-lei-ma} and hinders the transferability of findings to real-world ADS, undermining their trustworthiness~\cite{2023-Stocco-TSE}.

Various strategies have been proposed to mitigate the reality gap. Some aim to increase simulation fidelity through high-precision modeling (digital twins~\cite{DTwins,arcaini-foundation-models-digital-twins,lu2025foundationmodelssoftwareengineering}), others address specific gap dimensions, such as the perception gap, by translating simulated sensor outputs into more realistic versions using generative models~\cite{amini-gap,lambertenghi_ICST24,2023-Stocco-EMSE,2023-Stocco-TSE,10880098}. However, these methods are limited to model-level testing~\cite{2020-Haq-ICST,Codevilla} and do not capture the system-level interactions between perception, planning, and control modules that govern vehicle motion. As a result, they are susceptible to actuation gaps and often miss critical system-level failures~\cite{2020-Haq-ICST,2023-Stocco-EMSE,briand-offline-emse}.
Other strategies involve vehicle-in-the-loop (ViL) and mixed reality (MR) testing~\cite{HIL_survey}, by integrating physical components such as ECUs, small-scale robots, or full vehicles into simulation loops. While ViL provides closed-loop evaluation, it remains partially virtualized and fails to capture real-world imperfections, such as sensor noise and lighting variability (perception gap)~\cite{lambertenghi_ICST25}. MR partially mitigates this by injecting virtual elements (e.g., obstacles) into real sensor data, enriching scenario realism. 
Prior system-level studies using small-scale robots investigating failure transferability~\cite{2023-Stocco-TSE,9412011,viitala,DBLP:journals/corr/abs-1909-03467,DBLP:journals/corr/abs-1911-01562,khatiri2023simulation} have primarily documented the existence of the reality gap, without isolating their root causes or comparing how different test modalities influence gap reduction.

To this aim, in this paper, we conduct an empirical study of the reality gap in autonomous driving by comparing four representative testing modalities: SiL, ViL, MR, and full real-world execution (RW). Our goal is to characterize the dimensions of the reality gap, gap-specifically along actuation, perception, and behavioral fidelity, and to assess the degree to which each testing setup retains ADS behavior relative to real-world ground truth behavior. 
While prior work has evaluated or mitigated specific aspects of the reality gap, a broader evaluation spanning multiple testing modalities and ADS architectures remains unaddressed.

To investigate the impact of different testing modalities, we implemented a modular ROS-based evaluation framework that supports direct comparisons across synthetic, hybrid, and physical testing conditions. Our setup integrates both modular and end-to-end ADS architectures on a small-scale platform equipped with real sensors (camera and LiDAR) and its digital twin. 
We conduct \changed{hundreds} of tests across matched driving scenarios with shared road layouts, obstacle placements, and environmental conditions, allowing direct attribution of performance differences to the testing modality. 

Our findings reveal that:
(i)~SiL underestimates real-world variability due to idealized dynamics;
(ii)~ViL improves actuation realism but retains perception limitations;
(iii)~MR offers the best perceptual fidelity by blending virtual elements into real sensor data.
By isolating the effects of actuation, perception, and behavior on the reality gap, we find that critical failures often manifest differently across configurations, with the perception gap playing a greater role in behavioral divergence than actuation discrepancies. This underscores the importance of testing methods that retain real-world sensor complexity. Our results reveal the limitations of conventional simulation and point to MR as a practical middle ground between fidelity and scalability. 

Our paper makes the following contributions:

\begin{description}[noitemsep]
\item[Evaluation Framework.]
We provide a ROS framework for comparing SiL, ViL, MR, and RW testing for E2E and modular ADS, which is available~\cite{replication-package}.
\item[Empirical Study of the Reality Gap.]

We present a systematic analysis of the reality gap in ADS testing across behavior, actuation, and perception fidelity, isolating which failures transfer across test modalities. We show that MR testing uniquely replicates real-world system failures, outperforming SiL and ViL across all metrics.
\end{description}

\section{Background}\label{sec:background}

\subsection{Autonomous Driving Systems}\label{sec:ads}

Architecturally, ADS can be divided into two categories: end-to-end (E2E) systems and modular systems. E2E systems rely on deep neural networks (DNNs) that directly map camera inputs to driving commands such as steering, throttle, and braking. Once trained, models like NVIDIA's \mbox{Dave-2}~\cite{nvidia-dave2} or InterFuser~\cite{interfuser} infer vehicle control actions from raw sensor data without intermediate representations.
In contrast, modular ADS architectures such as Pylot~\cite{pylot}, Transfuser~\cite{Chitta2023PAMI}, and Autoware~\cite{kato} decompose the driving task into distinct components such as perception, planning, and control~\cite{yurtsever2020survey}. The perception module processes raw sensor data (e.g., LiDAR) to detect relevant objects and position them in the perceived environment map. The planning module uses this map to select a safe and feasible route, which the control module executes through low-level actuation commands.
As both architectures are actively used and researched~\cite{yurtsever2020survey,chen2023endtoend}, we include both in our evaluation to ensure broader applicability of our findings.

\begin{figure}[t]
\centering
   \includegraphics[width=0.9\columnwidth,trim={0.25cm 0.95cm 0.21cm 0.21cm},clip]{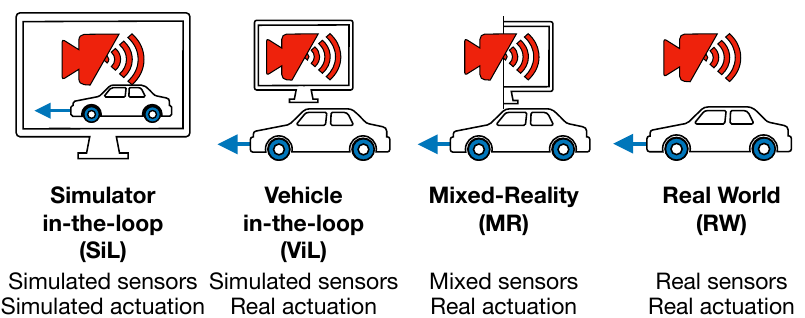}  
\caption{System-level testing modalities for ADS.}
\label{fig:testing-levels}
\end{figure}

\subsection{Reality Gap Dimensions in ADS Testing}

\autoref{fig:testing-levels} depicts the various system-level testing modalities for ADS. Real-world (RW) testing, conducted on closed-loop tracks or public roads, remains the gold standard for final validation. It exposes the ADS to real-life conditions~\cite{waymo-driver,Cerf:2018:CSC:3181977.3177753}, but is expensive, logistically complex, and time-consuming.

To support earlier development stages, simulation-based testing (SiL) offers a scalable and safe environment for experimenting across diverse scenarios. However, simulation introduces a reality gap, a mismatch between simulated and real-world behavior, largely due to limitations in replicating physical sensing and actuation with high fidelity.\footnote{While the term reality gap is also used in the literature to refer to the realism or real-world likelihood of test scenarios in scenario-based testing~\cite{reality-bites}, this aspect is beyond the scope of our work.} 
We refer to the former as the \textit{perception gap}, i.e., the inability of simulated sensors to accurately replicate the real-world sensors. The latter, the \textit{actuation gap}, reflects discrepancies between the vehicle dynamics modeled in simulation and those exhibited by physical vehicles. Together, these issues contribute to the \textit{behavior gap}, where the actions of the ADS in simulation diverge from its behavior in real-world scenarios.

To reduce this gap, vehicle-in-the-loop (ViL) methods embed a real vehicle into a simulated environment, enabling realistic actuation and closed-loop evaluation. While ViL helps address the actuation gap, it typically relies on synthetic sensor inputs and thus remains vulnerable to perception inaccuracies~\cite{amini-gap,lambertenghi_ICST24}.
Mixed reality (MR) testing extends ViL by blending simulated elements directly into real-world sensor streams (e.g., camera images, LiDAR point clouds). This approach preserves the physical characteristics of sensor signals and vehicle dynamics, aiming to jointly mitigate both the perception and actuation gaps.
\section{Reality Gap Evaluation Framework for ADS}

To date, a systematic assessment of the various dimensions of the reality gap, or the relative importance of the mitigation strategies, is missing, possibly due to the lack of a standardized evaluation framework. To address this, in this section, we introduce a framework designed to evaluate the transferability of system-level tests across different execution modalities, including SiL, ViL, MR, and real-world (RW) closed-loop testing. Our evaluation targets both E2E and modular ADS configurations for lane-keeping and obstacle-avoidance tasks. 

\subsection{Real-World Setup}\label{sec:real}

\subsubsection{Small-Scale Vehicle}

For our RW experiments, we use a small-scale vehicle based on the Donkey Car\textsuperscript{\texttrademark} open-source framework~\cite{donkeycar}, a widely adopted testbed for ADS research in both simulation and field settings~\cite{li2024autonomousdrivingsmallscalecars,CALEFFI2024271,Mokhtarian}. 
The vehicle is equipped with a front-facing 8MP Sony IMX219 RGB camera and a Time-of-Flight (ToF) sensor providing LiDAR-based depth at $256 \times 192$ resolution and up to 5~m range.

\subsubsection{Testing Tracks}\label{sec:tracks}

We conduct experiments in two indoor environments. 
\textit{Room Nominal} is a dedicated $6 \times 6$~m robotics lab with minimal background objects, providing a controlled setting. 
Its $4 \times 4$~m closed-loop track includes five right and two left curves ($10^\circ$–$90^\circ$), marked by 10~cm white lane margins and a central dotted line. 
In contrast, \textit{Room Generalization} is a larger, triangular multi-purpose room ($20 \times 10 \times 10$~m) with visually complex backgrounds. 
Its $6 \times 5$~m stadium-shaped track has two semi-circular curves connected by straights, bounded by narrower 3~cm lane markings and no central line. 
The floor color also differs, adding perceptual variability.

\subsubsection{Tracking system}

The tracking module determines the position of the vehicle and obstacles within the environment using a high-precision motion capture system. It utilizes four Vicon Bonita 10 cameras placed at the corners of each track with an extra margin of 1 m (both for Room Nominal and Generalization), which emit infrared light and detect reflections from retro-reflective markers attached to the vehicle and obstacles. Each object is uniquely identified based on the spatial configuration of its markers at 100 Hz. The Vicon tracker software manages the cameras, computes object poses, and broadcasts them as UDP messages. These messages are received by our framework, which integrates the data into the rest of the system.

\subsection{Software setup}

\subsubsection{Simulator}\label{sec:sim-testing}

The Donkey Car framework provides a high-fidelity Unity3D simulator that models the physical vehicle using the Nvidia PhysX engine. 
We extended the simulator by developing a new Unity scene that procedurally generates track layouts using Catmull-Rom splines~\cite{2025-Ali-ICSEW}, based on real-world lane margin data from Room Nominal and Generalization. To replicate the physical environments, we applied high-resolution (48 MP) floor images as textures and configured the lane markings to match each sandbox. 
In addition, we modified the simulator to support a virtual depth sensor that mimics the behavior of the ToF system used in our real-world setup. The virtual sensor produces depth data at the same resolution and provides intrinsic matrix parameters consistent with those retrieved by the ToF system. 

\begin{figure}[t]
\centering
\includegraphics[width=\linewidth]{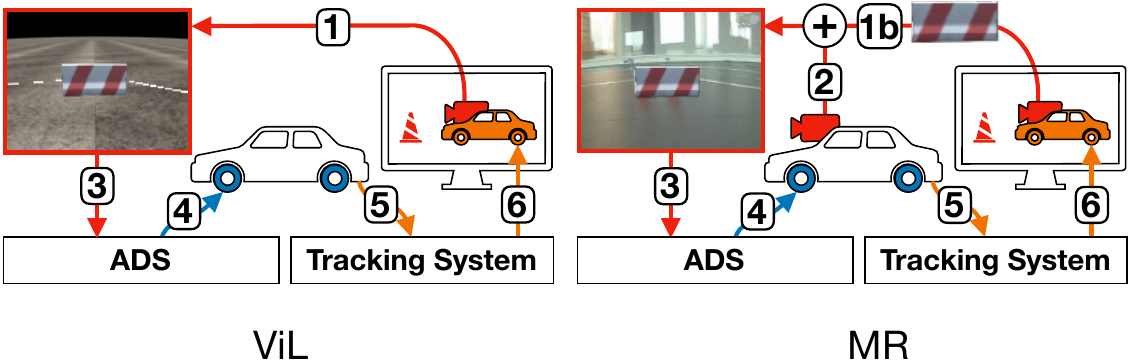}
\caption{Data flow and processing steps for ViL and MR.}
\label{fig:framework}
\end{figure}

\subsubsection{Testing Framework}\label{sec:vil-mr}

To support SiL, ViL, MR, and RW experimentation in a consistent setup, we integrate the simulator and physical platform into a ROS-based software framework~\cite{ros}.
ROS is chosen for its modular, platform-agnostic architecture, which allows the framework to be adapted across different hardware platforms, ADS designs, and simulation tools without major structural changes.
Core functionalities are implemented in dedicated ROS nodes. These nodes communicate via standard topics, enabling modular configuration and straightforward component replacement.

In the SiL modality, both the sensor stream and vehicle dynamics are handled entirely in simulation. The simulator provides perception data to the ADS, and control outputs are executed directly within the game engine, with no involvement of physical hardware or tracking. The framework simply interfaces the simulator with the ADS, routing image data in and receiving control commands back, and receiving telemetry data for modular positioning and experiments monitoring.

In the RW modality, perception data originates exclusively from the physical sensors onboard the vehicle. No simulated data is used. The tracking system is used only to monitor the vehicle's position and does not interact with the simulator. The framework sends the real sensor data to the ADS and routes the predicted driving commands to the vehicle.

For the ViL and MR modalities, \autoref{fig:framework} shows how the framework manages data flow between the simulator, the physical vehicle, and the ADS. Simulated data is first generated by the simulator. In the ViL setting, the full simulated sensor stream (e.g., camera or LiDAR) is used as input~\ding{182}. In the MR setting, only selected features or objects are rendered and extracted~\ding{182}(b), then merged with real sensor data from the physical vehicle~\ding{183} to form mixed-reality inputs.
The resulting sensor data, fully simulated in ViL or blended in MR-is processed by the ADS~\ding{184}, which outputs control commands for the physical vehicle~\ding{185}. The vehicle's motion is tracked in real time~\ding{186}, and its pose is fed back to the simulator~\ding{187} to maintain alignment between virtual and real environments. This forms a closed-loop system in which the ADS continuously perceives and acts on synchronized sensor data and vehicle dynamics.

In the rest of this section, we describe the main functionalities necessary to enable these testing procedures.

\noindent
\head{Control Modules}
For all testing levels, when using a modular ADS, the waypoint and speed commands produced by the ADS must be converted into actuation primitives. This is handled by two interfaces.
The \textit{Waypoint Follower} computes steering and throttle commands to reach target $(x, y)$ waypoints using a pure pursuit algorithm. A brake command is issued when the target is reached. It outputs commands in the format \texttt{[throttle, steering, brake]}.
The \textit{PID Speed Controller} receives the target, current speed, and the control commands. It modifies the throttle command component to maintain the target speed.

\noindent
\head{Simulator Interface} 
This component can retrieve rendered sensor outputs such as RGB and depth images. When testing in SiL, it accepts control commands (throttle, steering, brake) and optionally a throttle multiplier to drive the simulated vehicle. Finally, the interface publishes feedback (pose, velocity, obstacle positions) to be used by ADSs that require localization or monitoring behavior.

\head{Tracking Interface} 
This interface continuously tracks the vehicle and any physical obstacles using the external motion capture system. It publishes the vehicle pose, obstacle array, and vehicle speed estimated from pose deltas. This data is used by the modular ADS for localization, by the control modules for actuation, for synchronizing simulation, for evaluating control accuracy, and to provide ground-truth information.

\noindent
\head{Sensors Interfaces} 
These components stream raw sensor data from the physical vehicle. The camera interface provides RGB frames, while the LiDAR interface streams single-channel depth maps and publishes the intrinsic camera matrix for depth-to-3D conversion used in the perception pipeline.

\noindent
\head{LiDAR Pointcloud Generator}
This node converts depth maps from the ToF sensor into 3D point clouds using a standard pinhole camera projection model. Each pixel in the depth image is reprojected into 3D space according to:
\[
x = \frac{(i - c_x) \cdot z}{f_x}, \quad y = \frac{(j - c_y) \cdot z}{f_y}, \quad z = \text{depth}(i,j)
\]
where $(f_x, f_y)$ are the focal lengths and $(c_x, c_y)$ the principal point from the sensor's intrinsic matrix, which is published every frame by the sensor interface. Depth values are scaled and filtered using a minimum range threshold to suppress noise, and valid points are compiled into a point cloud. 
Transformations are published from the vehicle pose, whether from simulation or a real platform, to the LiDAR center frame, which is aligned with the actual center of the ToF sensor. This ensures spatial consistency of the projected point cloud within the vehicle's coordinate system.

\begin{figure}[t]
\centering
\includegraphics[width=0.8\columnwidth,trim={0 0.1cm 0 0},clip] {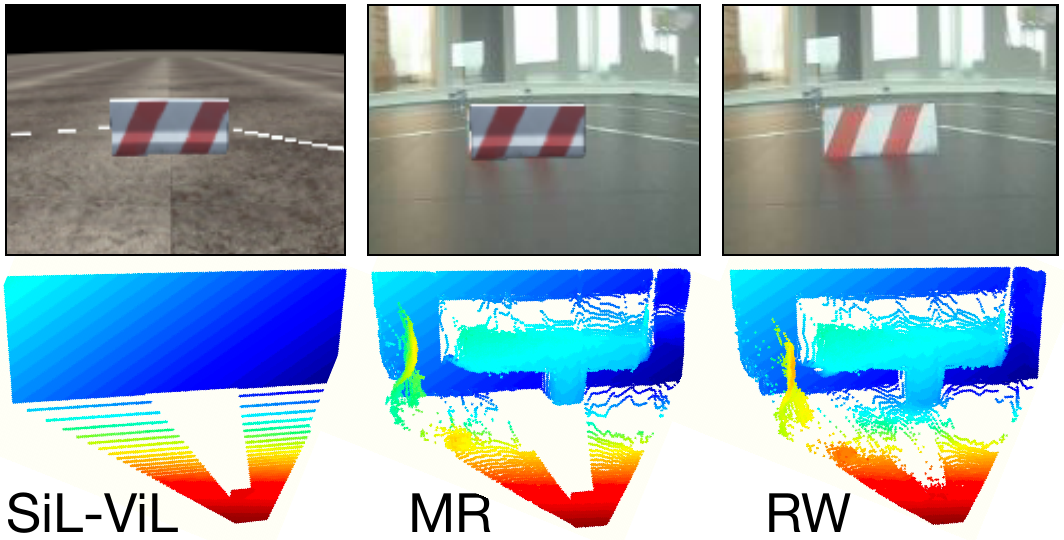}
\caption{Perception inputs across testing modalities (\changed{SiL-}ViL, MR, RW) for an obstacle positioned \changed{120 cm in front of the vehicle, using} the camera (top) and LiDAR (bottom) sensors.}
\label{fig:perc_examples}
\end{figure}

\subsubsection{Vehicle-in-the-Loop}\label{sec:vil}

To enable ViL, we extended the simulator interface to accept and apply the real vehicle's pose in simulation. This pose, received from the Tracking Interface, is used to update the state of the simulated vehicle, ensuring spatial alignment between the virtual and physical worlds.
In addition to the vehicle's pose, the simulator can receive the positions of tracked physical obstacles, each identified by a label and rendered in simulation using corresponding 3D models. This mirroring of real-world elements within the virtual environment is especially useful for maintaining consistent scenarios with RW for controlled experimental comparisons.
\subsubsection{Mixed-Reality}\label{sec:mr}

To enable MR, we extended the simulator interface to produce sensor outputs that render only certain components of the scene, in line with the approach by Shen et al.~\cite{simonwheels}. In this way, we can inject these rendered objects into the real sensor input. As for ViL, the vehicle is mapped to the real-world pose, and in this case, this adds the benefit of having the same sensor position in simulation and real world, so that merging data does not require spatial transforms.

\noindent
\head{Sensor Mixing Nodes}
These nodes produce mixed-reality inputs by blending simulated data into real sensor streams. For RGB, the node subscribes to both real camera feeds and simulator-generated RGBA images. Using the alpha channel as a per-pixel transparency mask, simulated RGB content is composited over the real feed via alpha blending. This yields a coherent visual stream that embeds virtual elements while preserving real-world context.
For depth, the node fuses grayscale depth images from real LiDAR and the simulator. At each pixel, the closer of the two depth values is selected, assuming virtual objects may occlude real ones.

Both pipelines run at 50 Hz and publish image messages that mimic native sensor output but include simulated features. \autoref{fig:perc_examples} illustrates an example of sensor-mixing on SiL/ViL, MR, and RW outputs, demonstrating spatial consistency across sensor types.

\subsubsection{\changed{Inference Location}}
\changed{In both ViL and MR, sensor data are generated directly on the workstation; to minimize latency, we run E2E inference and the modular pipeline locally. For fairness, inference in SiL and RW is also executed on the workstation, ensuring a uniform environment and comparable latency across domains.
}

\section{Empirical Evaluation}

\subsection{Research Questions}\label{sec:rqs}

\noindent
\textbf{RQ\textsubscript{1} (behavior gap):} \textit{How large is the behavior gap between SiL, ViL, MR, and real-world system testing?}

\noindent
\textbf{RQ\textsubscript{2} (actuation gap):} \textit{How large is the actuation gap between SiL, ViL, and \changed{RW} system testing?}

\noindent
\textbf{RQ\textsubscript{3} (input alignment and perception gap):} \textit{Are MR inputs aligned with RW? How much do they reduce the perception gap between ViL and \changed{RW} system testing?}

In \textbf{RQ\textsubscript{1}}, we aim to quantify the overall behavioral gap by comparing two representative ADS architectures across simulated (SiL), hybrid (ViL and MR), and \changed{RW} modalities. 

\textbf{RQ\textsubscript{2}} focuses on the actuation gap by comparing real and simulated vehicle responses to identical control inputs, focusing on the expected gap reduction thanks to the ViL modality. 

In \textbf{RQ\textsubscript{3}}, we investigate the perception gap by evaluating perceptual fidelity to the real world of the MR modality, focusing on the alignment and realism between original and sensor-mixed data and, consequently, the expected gap reduction thanks to the MR modality. 

\subsection{Testing Scenarios}

\begin{figure}[t]
\centering
\includegraphics[width=0.6\columnwidth]{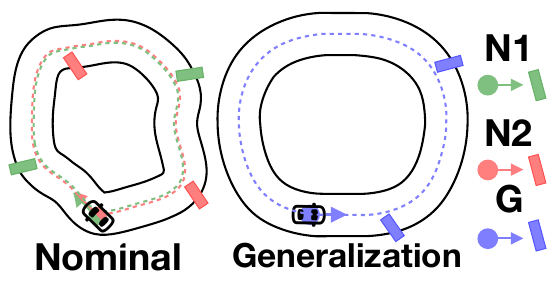}
\caption{Testing scenarios used to answer RQ\textsubscript{1}.}
\label{fig:testing_tracks}
\end{figure}

We define three evaluation scenarios across the two physical environments described in \autoref{sec:tracks}. 
A visual representation is shown in \autoref{fig:testing_tracks}.
In Room Nominal, we create two testing scenarios. \changed{As creating multiple real-world tracks manually is impractical, we design a single layout for N1 and N2 with turns reaching the vehicle’s steering limit, verified via the waypoint follower, and additional curves of varying curvature.} 
In Scenario N1, two static obstacles are placed on straight segments shortly after turns, providing clear visibility and sufficient reaction time. This \changed{low complexity} setup serves as a baseline for evaluating basic \changed{lane keeping} and obstacle avoidance. 
Scenario N2 increases difficulty by placing one obstacle directly on a turn and another immediately after it, within the vehicle's expected path. This configuration reduces visibility windows and imposes tighter spatial constraints, challenging both perception and planning.

In Room Generalization, \changed{for generalization (G), we reuse a printed track featuring turns and straight sections absent from the nominal layout. This new location introduces a perception shift by exposing the camera to unseen background regions.} 
One obstacle is placed on a straight segment (fully visible) and another on a curve (partially occluded). 
This setup is designed to test the generalization of reality gap mitigation techniques (ViL and MR) on an unseen environment, visual characteristics, and object placements.

\subsection{Automated Driving Systems}\label{sec:sut}

\subsubsection{End-to-End}

The end-to-end ADS is a Dave-2~\cite{nvidia-dave2} imitation learner trained on real-world camera data to follow the center lane and shift laterally to avoid obstacles with a 0.8m buffer. We recorded 50 laps in Room Nominal at 50 FPS with varying obstacle placements, applied horizontal flipping, and obtained nearly 180k samples. Some laps included recovery from off-track behavior.
The training was performed using existing guidelines~\cite{nvidia-dave2}, with a batch size of 64, and a learning rate of 0.0001 for 500 epochs, and the Adam~\cite{kingma2014adam} optimizer to minimize the MSE loss, with early stopping (patience 30, $\Delta$ MSE $<$  0.05). 
The final model achieved 0.02 MSE. 

\subsubsection{Modular}

The modular ADS features a LiDAR-based perception module, a lattice planner, and a control stack with waypoint following and PID speed regulation. This architecture reflects the modular principles adopted in real-world stacks such as Apollo~\cite{apollo} and Autoware~\cite{kato}, albeit simplified for a small-scale testbed. In particular, the perception module adopts a LiDAR  DBSCAN clustering approach for obstacle detection\cite{dbscan}, a technique widely used in autonomous driving systems~\cite{modularlidar1,modularlidar2,modularlidar3}. The planning module uses a Fr\'enet-frame lattice planner, in line with established motion planning methods employed in research and industrial pipelines~\cite{modularplanner1,modularplanner2,modularplanner3}.

\changed{\subsection{Experimental Setup Validation}
For ViL and MR, the real vehicle’s motion must be continuously mapped into the simulation, requiring real-time execution to keep the simulated state and generated sensor data synchronized with the physical vehicle. Across all domains, the E2E or modular pipeline must execute fast enough to deliver control commands (simulated or real) without latency, as delays could distort the observed behavior.}
\changed{To verify this, we benchmarked the framework over 5000 samples, measuring execution time for every component. All modalities (SiL, ViL, MR, RW) ran in real time: the simulator achieved $\approx 80$ FPS ($\approx 12.5$ ms per frame) on our workstation (AMD Ryzen 5, 16 GB RAM, NVIDIA RTX 2060). The tracking system has been measured (at the simulator-level) to provide state updates at $\approx 99.93$ Hz ($\approx 10.01$ ms). Since the onboard camera and LiDAR sensors operate at 20 FPS (50 ms per frame), we configure the simulator to run its camera at 20 FPS, and to avoid unnecessary system load, we cap the global simulation frame-rate at 60 FPS.
}
\changed{The E2E ADS executed in $\approx 15$ ms on average, and the modular ADAS in $\approx 20$ ms. Inference begins as soon as a new sensor message arrives and can run in parallel if needed, though both pipelines consistently complete before the next frame. Control commands, measured at the vehicle-level, were issued every $\approx 50.5$ ms, remaining stable across all experiments.}

\changed{To avoid packet loss, we use a dedicated TP-Link Archer~C6 dual-band gigabit WLAN router (867~Mbit/s 5~GHz + 300~Mbit/s 2.4~GHz, 4~Gigabit LAN ports) configured as a local-only network. 
Vehicle–workstation communication uses TCP over the 5~GHz link (Intel-8265AC, 867~Mbit/s), while tracking–workstation communication is performed via wired Gigabit Ethernet. 
JPEG-compressed camera images ($\approx800$~kB) and compact control commands ($<30$~bytes) were observed to fit within each 20~Hz cycle, allowing $\approx50$ image transmissions and thousands of control messages without observed packet loss.}

\changed{We also validated the accuracy of the tracking system using a calibrated reference object with five retro-reflective markers mounted in a fixed, non-coplanar geometry. 
The employed Vicon system performs tracking by matching such known geometries to objects in space, allowing it to estimate a unique 6-DoF pose (position and three-axis orientation) for each tracked object. 
To verify accuracy, we measured the distance between two markers of the calibration object with a known spacing of 240\,mm, repeating the measurement five times at 30\,s intervals in both Room Nominal and Room Generalization. 
The mean measured distance was 240 mm, with a mean absolute error of 0.40 mm and a standard deviation of 0.0253 mm, confirming sub-millimeter position accuracy. 
Because orientation is derived from the same geometry matching process, a correct match of the marker configuration also guarantees accurate three-axis orientation. 
These results demonstrate accurate vehicle localization and synchronization for ViL and MR experiments.
}

\subsection{Procedure and Metrics}

\subsubsection{RQ\textsubscript{1} (behaviour gap)}

To address \textbf{RQ\textsubscript{1}}, we evaluate both ADS architectures (E2E and modular) across four domain configurations: the real world (ground truth), SiL (baseline), ViL, and MR. Experiments are conducted for scenarios N1, N2 (for nominal), and G (for generalization).

\changed{In the nominal setting, for each ADS, we run the RW testing modality on scenarios N1 and N2 at least five times and continue until we obtain at least four successful executions. Then, for the SiL, ViL, and MR testing modalities, we run the same number of tests as in RW and quantified the differences in trajectory fidelity, task completion, lane-keeping, and obstacle avoidance.} More in detail, we assess system behavior using two categories of metrics. First, we measure trajectory similarity using the Fr\'echet distance~\cite{cheng2024frechetdistancesubquadratictime} between each run and its real-world counterpart. 
\changed{The Fr\'echet distance accounts for both spatial proximity and the order of points along the trajectory, making it well-suited for autonomous driving where both path shape and sequence matter. 
We compute it between each SiL, ViL, or MR run and the corresponding RW run to quantify trajectory deviation. 
This metric has been widely adopted in prior work~\cite{cheng2024frechetdistancesubquadratictime,lambertenghi_ICST24} to assess behavioral similarity in autonomous systems.}
Second, we evaluate task performance through several sub-metrics. The \textit{failure rate} is defined as the proportion of runs resulting in either a collision with an obstacle or a lane departure. We also report the absolute number of \textit{obstacle crashes} and \textit{out-of-road} events. To characterize failures, we compute the \textit{completion rate} as the percentage of the track driven before a failure. \changed{For SiL, ViL, and MR, the \textit{completion rate} is reported relative to the RW execution.}
\changed{This setting resulted in 92 executions.}

In Scenario G, we only evaluate RW, SiL, and MR modalities, omitting ViL, as the room's features are expected to mainly affect perception. Each configuration is executed three times, and the same set of metrics, trajectory similarity, failure rate, completion rate, and driving quality, are used to assess behavioral consistency across domains.
\changed{This setting results in 18 executions.}

Finally, for the modular ADS, which separates perception and control, we conduct an ablation study. In this configuration, we bypass the perception module and provide ground-truth obstacle locations directly to the planner in both the SiL and RW modalities. This isolates the impact of actuation by removing the perception gap, an analysis not possible for the E2E system, where perception and control are coupled together.
\changed{By re-purposing the results of the RW and SiL models in scenarios N1 and N2, this setting results in 16 new executions (8 runs x 2 domains).}

\subsubsection{RQ\textsubscript{2} (actuation gap)}

To address \textbf{RQ\textsubscript{2}}, we quantify the actuation gap by comparing the physical response of the real vehicle to that of a simulated vehicle under identical control inputs.
We limit the evaluation to SiL, ViL, and RW, omitting MR as this modality does not change actuation fidelity beyond what is already captured by ViL. \changed{Each experiment was repeated five times to capture variability, and data were logged at 100 Hz, yielding large frame-level sample sizes. Statistical analysis is applied only to per-frame metrics, where sample sizes support meaningful significance testing.}
We design five test scenarios spanning both low-level actuation and high-level control, described next.
\noindent
\head{Forward Motion} 
Constant throttle is applied with steering fixed at zero and braking disabled. 
\changed{Throttle commands are taken directly from the RW-recorded ROS messages with matched timestamps.}
Trials run for three seconds using throttle values of 0.34, 0.365, and 0.39, which represent the lowest throttle at which the vehicle starts moving, to the highest throttle at which the vehicle can experiment while remaining in the tracking area. 
We compute the total distance traveled, average speed, and trajectory, and compare both speed and trajectory distributions.
\changed{This setup results in 15 executions (3 speed values × 5 repetitions).}

\noindent
\head{Steering Motion}
With the throttle fixed at 0.365 and braking disabled, selected based on the medium throttle value identified in \textit{Forward Motion experiments}.
Steering commands are issued across six values: $\{-1.0, -0.6, -0.3, 0.3, 0.6, 1.0\}$, representing the full steering range, from full left to full right. 
\changed{As with throttle, steering inputs are replayed from recorded ROS messages to ensure the same timing across domains.}
We compare turning radius via circle fitting, raw trajectories, and trajectory distributions.
\changed{This setup results in 30 executions (6 steering values × 5 repetitions).}

\noindent
\head{Braking Motion}
Following forward motion, braking is triggered 35 cm before a 2.0 m goal. Steering remains at zero, and throttle inputs match those from the forward motion test. 
\changed{Braking events are triggered based on each domain’s local pose estimate, so that stopping occurs at the same physical location in RW, SiL, and ViL.
}
We measure braking distance, approach speed, and deceleration. 
\changed{This setup results in 15 executions (3 throttle values x 5 repetitions).}

\noindent
\head{PID Speed Control}
This scenario evaluates closed-loop speed regulation. The throttle is fixed at 1.0 while a PID controller adjusts output via a throttle multiplier.
\changed{We replay the exact RW speed requests for each phase to maintain identical target profiles across domains.}
The vehicle drives in a circular path (steering at 0.6) across four 10-second speed phases: 0.4, 0.8, 0.6, and 0\,m/s. We report comparisons of per-phase and overall speed errors. Trajectory similarity is not computed due to the fixed path.
\changed{This setup results in 5 executions.}

\noindent
\head{Waypoint Following}
We test tracking performance on predefined waypoint paths (throttle fixed at 0.365), ranging from simple (single point) to complex (wide turns, sharp turns, S-shapes).
\changed{The same actuation module and waypoint-following logic are used in all domains, running closed-loop control on local pose estimates. 
All waypoint-following tests were performed before adding lane markings in Nominal, as perception is not involved in this experiment.}
\changed{We compute} Fr\'echet distances among trajectories.
\changed{This setup results in 30 executions (6 paths × 5 repetitions).}

\subsubsection{RQ\textsubscript{3} (perception validity and gap)}

To evaluate \textit{perception input validity}, we used two test scenarios. All experiments are repeated five times from a fixed initial pose in each domain to capture variability.

\noindent
\head{Obstacle Placement}
Static obstacles are placed at 0.4–1.6 m distances, either centrally (single) or symmetrically (dual) within the sensor FoV. 
For cameras, we compute IoU between manually annotated real and mixed bounding boxes (10 samples/modality). For LiDAR, we calculate the Euclidean distance between cluster centroids. \changed{This experiment consists of 40 executions (8 obstacle configurations × 5 repetitions)}

\noindent
\head{Lane Placement}
Using a real trajectory on Room Nominal, lane overlays are rendered in mixed reality with partial transparency. The vehicle runs in both directions with five lateral behaviors (center, margins, halves). \changed{The data is collected five times per configuration, to capture variability.} 
An independent human annotator marked five vertical anchor points per image; Catmull-Rom splines are fitted, and alignment is evaluated via Euclidean distance between 100 sampled points per spline. \changed{This experiment consists of 50 vehicle runs (5 lane displacements × 2 directions × 5 repetitions)}.

\begin{table*}[t]
\caption{RQ\textsubscript{1}: Behavior gap results. \changed{Trajectory $\Delta$ reports the Fr\'echet distance between the trajectory in each domain and the RW trajectory (lower is better).
Completion\%$\Delta$ is relative to RW, where negative/higher values indicate lower/higher track completion than RW. FR = failure rate. Ablation obstacles: GT = ground truth, P = perception-detected.}}
\setlength{\tabcolsep}{0.8pt}
\label{tab:rq1}
        \centering
        \setlength{\tabcolsep}{2pt}
        \renewcommand{\arraystretch}{1.2}
        \resizebox{\textwidth}{!}{%
        \begin{tabular}{lcccccccc@{\hskip 1em}cccccccc@{\hskip 1em}cccccc@{\hskip 1em}cccc}
        
        \toprule
        & \multicolumn{16}{c}{Nominal} & \multicolumn{6}{c}{Generalization (G)} & \multicolumn{4}{c}{Ablation (N2)}\\
        \cmidrule(lr){2-17}
        \cmidrule(lr){18-23}
        \cmidrule(lr){24-27}
        & \multicolumn{8}{c}{E2E} & \multicolumn{8}{c}{Modular} & \multicolumn{3}{c}{E2E}& \multicolumn{3}{c}{Modular}& \multicolumn{4}{c}{Modular}\\
        \cmidrule(lr){2-9}
        \cmidrule(lr){10-17}
        \cmidrule(lr){18-20}
        \cmidrule(lr){21-23}
        \cmidrule(lr){24-27}
        & \multicolumn{2}{c}{\changed{RW}} & \multicolumn{2}{c}{\changed{SiL}} & \multicolumn{2}{c}{ViL} & \multicolumn{2}{c}{MR} & \multicolumn{2}{c}{\changed{RW}} & \multicolumn{2}{c}{\changed{SiL}} & \multicolumn{2}{c}{ViL} & \multicolumn{2}{c}{MR } & \changed{RW} & \changed{SiL} & MR & \changed{RW} & \changed{SiL} & MR & \multicolumn{2}{c}{\changed{RW}} & \multicolumn{2}{c}{\changed{SiL}}\\
        \cmidrule(lr){2-3}
        \cmidrule(lr){4-5}
        \cmidrule(lr){6-7}
        \cmidrule(lr){8-9}
        \cmidrule(lr){10-11}
        \cmidrule(lr){12-13}
        \cmidrule(lr){14-15}
        \cmidrule(lr){16-17}
        \cmidrule(lr){18-20}
        \cmidrule(lr){21-23}
        \cmidrule(lr){24-25}
        \cmidrule(lr){26-27}
         &N1&N2&N1&N2&N1&N2&N1&N2&N1&N2&N1&N2&N1&N2&N1&N2&G&G&G&G&G&G&GT&P&GT&P\\
        \midrule
        Trajectory$\Delta$ & - & - & 0.53 & 3.25 & 0.43 & 3.55 & 0.27 & 0.30           & - & - & 0.31 & 1.37 & 0.32 & 1.49 & 0.19 & 0.84             & - & 3.73 & 0.40 & - & 0.37 & 0.15 & - & - & 0.25 & 1.37\\
        Completion\%$\Delta$ & 100 & 100 & 0 & -56 & 0 & -57 & 0 & 0 & 100 & 62 & 0 & +38 & 0 & +31 & 0 & +13 & 37 & +63 & 0 & 100 & 0 & 0 & 100 & 62 & 0 & +38\\
        \# Offroad & 0 & 0 & 0 & 5 & 0 & 5 & 0 & 0                                & 0 & 2 & 0 & 0 & 0 & 1 & 0 & 2                             & 3 & 0 & 3 & 0 & 0 & 0 & 0 & 2 & 0 & 0 \\
        \# Crashes & 0 & 0 & 0 & 0 & 0 & 0 & 0 & 0                                    & 0 & 2 & 0 & 0 & 0 & 1 & 0 & 2                             & 0 & 0 & 0 & 0 & 0 & 0 & 0 & 2 & 0 & 0 \\
        FR &  0/5 & 0/5  &  0/5 & 5/5  &  0/5 & 5/5  &  0/5 & 0/5                           & 0/5 & 4/8 & 0/5 & \changed{0/8} & 0/5 & \changed{2/8} & 0/5  & \changed{4/8}                        & 3/3 & 0/3 & 3/3 & 0/3 & 0/3 & 0/3 & 0/8 & 4/8 & 0/8 & 0/8 \\      
        \bottomrule
        \end{tabular}
        }
\end{table*}

Concerning the \textit{perception gap}, we compare sensor outputs from ViL, MR, and RW runs using pixel-level and geometric similarity metrics. We exclude SiL from this analysis, as it uses the same simulated perception as ViL but does not include real-world vehicle and obstacle mapping. Our analysis uses 100 synchronized samples collected during the obstacle placement experiments, spanning 8 configuration variants for both camera and LiDAR (800 images or point clouds/domain).

\noindent
\head{Camera}
We apply twelve image similarity metrics, identified from previous literature~\cite{lambertenghi_ICST24}, to assess realism and alignment: correlation coefficient, histogram intersection, Local Binary Pattern (LBP) histogram similarity, Peak Signal-to-Noise Ratio (PSNR), Structural Similarity Index Measure (SSIM), normalized mutual information (NMI), image fractal dimension (IFD), Kullback–Leibler (KL) divergence, Mean Squared Error (MSE), perceptual distance derived from deep features using VGG16\changed{\cite{Simonyan2014}}, texture similarity based on gray-level co-occurrence matrix (GLCM) properties, and Wasserstein distance (WD). These metrics collectively capture pixel-level, structural, and perceptual fidelity. Improved scores in MR relative to ViL would indicate perception gap reduction. 

\noindent
\head{LiDAR}
We compare 3D point clouds from the same matching poses as for the camera, across ViL, MR, and RW domains by computing the mean, maximum, and standard deviation of Euclidean distances between corresponding points.

\subsubsection{Summary}
Our experiments comprise over \changed{311} test executions: \changed{126} ADS runs evaluated with four behavioral metrics in RQ\textsubscript{1} (92 for N1/N2, 18 for G, and 16 for ablation), 95 actuation tests across three domains with an average of three metrics per domain (RQ\textsubscript{2}), and 90 perception runs assessed with 12 camera and 3 LiDAR metrics (RQ\textsubscript{3}). 
In total, the study covers more than 1.9 million ADS predictions and over 20 hours of RW execution. 
Including SiL, ViL, and MR domains, the total experimental runtime exceeds 185 hours.

\section{Results}

\subsection{RQ\textsubscript{1} (behaviour gap)}

\autoref{tab:rq1} shows the results for the behavior gap. For each scenario and testing modality, the table reports the trajector\changed{y's Frechet distance to RW (Trajectory$\Delta$)}, the \changed{difference} in completion rate \changed{compared to RW} \changed{(Completion\% $\Delta$)}, the number of out-of-road and crash episodes, and the failure rate \changed{(FR)}.
\changed{Reported results, with the exception of FR, are averaged across the multiple repeated executions of each configuration.}

The trajectory gap between SiL and \changed{RW} is especially pronounced for the E2E system in scenario N2, with a Frechet distance of 3.25m. In N1, the gap is smaller (0.53m). For the modular system, the gap is more modest across both scenarios: 0.31m (N1) and 1.37m (N2).
ViL has a mixed impact on trajectory alignment. For E2E, the gap reduces slightly to 0.43m in N1 but worsens to 3.55m in N2. For modular, the change is negligible-0.32m in N1 and 1.49m in N2.
In contrast, MR significantly improves trajectory alignment. For E2E, MR reduces the gap to 0.27m in N1 and \changed{0.30} in N2, achieving over 90\% reduction from SiL. For modular, MR reduces the gap to 0.19m in N1 and 0.84m in N2, representing a 39\% improvement over SiL.

\begin{figure}[t]
\centering
\includegraphics[width=0.85\columnwidth]{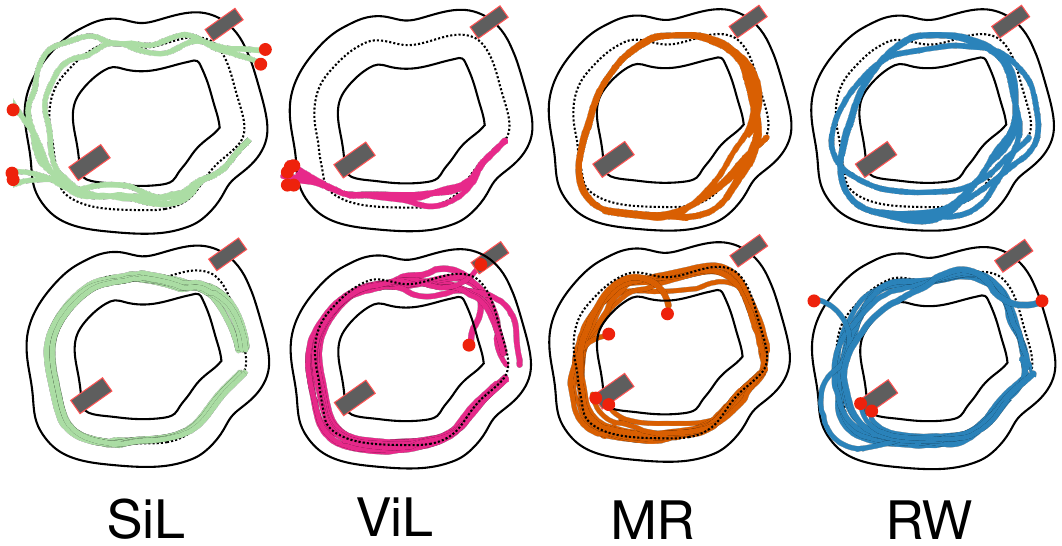}
\caption{Trajectory traces from the nominal experiment in scenario N2 (RQ\textsubscript{1}) for E2E (top) and modular ADS (bottom). \changed{Obstacles: red rectangles, failures: red dots.}}
\label{fig:example_rum}
\end{figure}

Task success reveals distinct patterns across scenarios. In N1, both ADS architectures succeed in all runs across all modalities.
In contrast, in scenario N2, the E2E system fails all SiL and ViL runs, due to off-road departures during obstacle avoidance. MR fully restores correct behavior, with no failures and a perfect completion rate, closing the gap.

The Modular system also completes all N1 runs successfully. In N2, the \changed{RW} configuration fails in 4 out of 8 runs, two due to off-road departures and two due to obstacle collisions. SiL, however, shows no failures, overestimating system robustness. ViL partially reintroduces realistic failures (2/8), while MR closely matches \changed{RW} outcomes with 4/8 failures of the same type and location (\autoref{fig:example_rum}).

For the E2E ADS, both SiL and ViL consistently fail during obstacle avoidance, resulting in lane departures. The plot makes clear that MR and RW produce highly similar trajectories, with no failures. For the Modular ADS, MR, and RW again show strong alignment, exhibiting obstacle collisions at the same location, though lane departures differ slightly in timing and position. In contrast, ViL fails at a different obstacle and shows a lane departure in a distinct area, highlighting its partial but inconsistent realism.

Concerning generalization, E2E fails in all three \changed{RW} runs (completion 0.37), while \changed{it} succeeds in all \changed{SiL runs}. This likely results from SiL's static visuals, which obscure real-world domain shifts. MR corrects this mismatch, producing 3/3 failures (matching \changed{RW}) and reducing trajectory error from 3.73 m (SiL) to 0.40 m.
The modular system performs robustly in G across \changed{RW}, SiL, and MR with 100\% success, highlighting its insensitivity to visual domain shift due to its LiDAR-based perception, as observed for scenario N1.

Finally, we isolate the cause of failure in modular N2 by bypassing perception and feeding ground-truth obstacle positions to the planner. With perception enabled \changed{(columns P)}, the \changed{RW} system fails in 4 out of 8 runs, while SiL shows none, overestimating robustness by masking perception errors. With ground-truth inputs \changed{(columns GT)}, both \changed{RW} and SiL succeed in all runs, confirming perception as the failure source and showing that SiL hides these issues due to ideal inputs. Notably, SiL's trajectory error also drops from 1.37 m to 0.25 m, closely matching \changed{RW} behavior, indicating that once perception is removed, actuation in SiL aligns well with reality.

\begin{tcolorbox}[boxrule=0pt,frame hidden,sharp corners,enhanced,borderline north={1pt}{0pt}{black},borderline south={1pt}{0pt}{black},boxsep=2pt,left=2pt,right=2pt,top=2.5pt,bottom=2pt]
\textbf{RQ\textsubscript{1}}: \textit{
Simulation-in-the-loop (SiL) often fails to reflect real-world \changed{(RW)} outcomes, producing both false positives and false negatives; Mixed-Reality (MR) consistently aligns best with \changed{RW} behavior across scenarios and architectures, reproducing real failures and improving trajectory fidelity.
}
\end{tcolorbox}

\subsection{RQ\textsubscript{2} (actuation gap)}

\begin{table}[t]
\caption{RQ\textsubscript{2}: Actuation gap results.}
\label{tab:rq2}
    \centering
    \footnotesize
    \setlength{\tabcolsep}{6.3pt}
    \renewcommand{\arraystretch}{1.1}
\begin{tabular}{lcccccc}
\toprule
& \multicolumn{2}{c}{Real} & \multicolumn{2}{c}{SiL$\Delta$} & \multicolumn{2}{c}{ViL$\Delta$}\\
\midrule
\multicolumn{7}{l}{\textbf{Throttle}}\\
 Distance travelled  &  \multicolumn{2}{c}{2.87}& \multicolumn{2}{c}{0.87 ± 0.23} & \multicolumn{2}{c}{\textbf{-0.01 ± 0.85}}\\
 Average speed & \multicolumn{2}{c}{0.61} & \multicolumn{2}{c}{-0.43 ± 0.01} & \multicolumn{2}{c}{\textbf{-0.07 ± 0.13}}\\
 Trajectory diff. & \multicolumn{2}{c}{-} & \multicolumn{2}{c}{0.25 ± 0.12} & \multicolumn{2}{c}{\textbf{0.001 ± 0.00}}\\
 Trajectory eff. size  & \multicolumn{2}{c}{-} & \multicolumn{2}{c}{\textcolor{gray}{-0.22}} & \multicolumn{2}{c}{\textcolor{gray}{0.00}}\\
\midrule
\textbf{Steering} & L & R & L& R & L & R\\
Radius & 1.79 & 1.14 & -0.81 & -0.16 & \textbf{0.00} & \textbf{0.00} \\
Radius (std) & 0.07 & -0.02 & -0.07 & +0.02 & \textbf{0} & \textbf{0} \\
Trajectory diff. & \multicolumn{2}{c}{-} & 0.58 & 0.15 & \textbf{0.002} & \textbf{0.007} \\
Trajectory eff. size & \multicolumn{2}{c}{-} & \underline{0.50} & \textcolor{gray}{-0.06} & \textcolor{gray}{\textbf{0.00}} & \textcolor{gray}{0.003} \\
\midrule
\multicolumn{7}{l}{\textbf{Braking}}\\
Braking dist. & \multicolumn{2}{c}{0.18} & \multicolumn{2}{c}{\textbf{-0.02 ± 0.06}} & \multicolumn{2}{c}{0.05 ± 0.02}\\
Speed         & \multicolumn{2}{c}{0.43} & \multicolumn{2}{c}{-0.14 ± 0.04} & \multicolumn{2}{c}{\textbf{-0.01 ± 0.003}}\\
Acceleration (e-2) & \multicolumn{2}{c}{-0.006} & \multicolumn{2}{c}{0.82 ± 0.06} & \multicolumn{2}{c}{\textbf{0.06 ± 0.07}}\\
Speed eff. size   & \multicolumn{2}{c}{-} & \multicolumn{2}{c}{\underline{0.514}} & \multicolumn{2}{c}{\textbf{\underline{ 0.054}}}\\
\midrule
\multicolumn{7}{l}{\textbf{PID control}}\\
Avg. Speed $\epsilon$ & \multicolumn{2}{c}{0.26} & \multicolumn{2}{c}{0.07 ± 0.24} & \multicolumn{2}{c}{\textbf{0.04 ± 0.23}}\\
T. Speed 0.4 $\epsilon$ & \multicolumn{2}{c}{0.38} & \multicolumn{2}{c}{\textbf{0.01 ± 0.004}} & \multicolumn{2}{c}{-0.05 ± 0.03}\\
T. Speed 0.8 $\epsilon$ & \multicolumn{2}{c}{0.45} & \multicolumn{2}{c}{0.00 ± 0.19} & \multicolumn{2}{c}{-0.00 ± 0.15}\\
T. Speed 0.6 $\epsilon$ & \multicolumn{2}{c}{0.14} & \multicolumn{2}{c}{-0.04 ± 0.05} & \multicolumn{2}{c}{\textbf{0.02 ± 0.07}}\\
T. Speed 0.0 $\epsilon$ & \multicolumn{2}{c}{0.23} & \multicolumn{2}{c}{0.14 ± 0.25} & \multicolumn{2}{c}{\textbf{0.06 ± 0.26}}\\
Speed eff. size & \multicolumn{2}{c}{-} & \multicolumn{2}{c}{\underline{-0.30}} & \multicolumn{2}{c}{\textbf{\underline{ -0.17}}}\\
\midrule
\multicolumn{7}{l}{\textbf{Waypoint control Trajectories}}\\
Straight & \multicolumn{2}{c}{-} & \multicolumn{2}{c}{0.07 ± 0.04} & \multicolumn{2}{c}{\textbf{0.008 ± 0.003}}\\
Close & \multicolumn{2}{c}{-} & \multicolumn{2}{c}{0.08 ± 0.03} & \multicolumn{2}{c}{\textbf{0.008 ± 0.001}}\\
Far & \multicolumn{2}{c}{-} & \multicolumn{2}{c}{0.11 ± 0.04} & \multicolumn{2}{c}{\textbf{0.01 ± 0.004}}\\
Sharp & \multicolumn{2}{c}{-} & \multicolumn{2}{c}{0.38 ± 0.02} & \multicolumn{2}{c}{\textbf{0.01 ± 0.005}}\\
Curve & \multicolumn{2}{c}{-} & \multicolumn{2}{c}{0.13 ± 0.05} & \multicolumn{2}{c}{\textbf{0.01 ± 0.004}}\\
\bottomrule
\end{tabular}
\end{table}

\autoref{tab:rq2} shows the actuation gap results across SiL, ViL, and real-world domains. 
For throttle and braking, we report the average across our three predefined throttle values, while for steering, we aggregate profiles with positive steering angles as right turns and negative angles as left turns. 
Across all tests, we assessed the statistical significance of the differences across modalities using the non-parametric Mann-Whitney U test~\cite{Wilcoxon1945} (with $\alpha = 0.05$) and the magnitude of the differences using Cohen's $d$ effect size~\cite{cohen1988statistical}. Statistical tests are presented by effect size, with an underline representing significance.

Concerning throttle, SiL overestimates the traveled distance by 30.3\% ($0.87$ m \changed{over} $2.87$ m) and underestimates speed error by 70.5\% ($-0.43$ m/s error over 0.61 m/s). 
ViL nearly eliminates the distance error ($-0.01$ m) and reduces the speed error to 11.5\% ($-0.07$ m/s). The Fr\'echet distance improves from $0.25$ in SiL to $0.001$ in ViL. 

Concerning steering, 
SiL underestimates the turning radius by –0.81~m for left turns (45.3\% of the real radius) and –0.16~m for right turns, while ViL eliminates both errors (0.00~m). 
The Fr\'echet distance drops from 0.58 (left) and 0.15 (right) in SiL to 0.002 and 0.007 in ViL. 
For the left turn, statistical tests confirm a significant deviation in SiL with a medium effect size ($d=0.50$), whereas no significant difference is found for ViL. 
For the right turn, the results are not statistically significant. 
Together with the smaller SiL error in right turns, this suggests asymmetric real-vehicle steering behavior: the digital twin, likely calibrated on right turns, replicates those dynamics accurately but fails to capture the left-turn deviation, underscoring the impact of unmodeled steering dynamics.

Concerning braking, 
SiL underestimates braking distance by 12.2\% ($-0.02$ m \changed{over} 0.18 m real) and introduces a deceleration error of $0.0082$ m/s\textsuperscript{2}. ViL slightly overshoots braking distance (+5.1 cm) but reduces deceleration error to $0.0006$ m/s\textsuperscript{2}. The speed profile similarity shows statistical significance for both cases, with medium ($d=0.514$) and negligible ($d=0.054$) effect sizes for SiL and ViL, respectively. The smaller effect size in ViL confirms improved alignment.

Concerning PID speed control, 
SiL shows an average tracking error of 0.074~m/s (28\% of the real 0.267~m/s), while ViL reduces this to 0.040~m/s (15\%). 
At individual control points, ViL consistently improves accuracy—for example, at 0.4~m/s, SiL yields a +0.01~m/s error versus –0.05~m/s for ViL. 
Statistical tests confirm significance, with small effect sizes for SiL ($d=0.30$) and ViL ($d=0.17$). 

Concerning waypoint following, 
SiL shows its largest deviation in sharp turns (0.385~m), with smaller deviations in close (0.08~m), far (0.11~m), curve (0.13~m), and straight (0.07~m) paths. 
ViL reduces deviations across all path types, down to 0.008~m on straights and 0.01~m on sharp turns, achieving over 90\% error reduction.

\begin{tcolorbox}[boxrule=0pt,frame hidden,sharp corners,enhanced,borderline north={1pt}{0pt}{black},borderline south={1pt}{0pt}{black},boxsep=2pt,left=2pt,right=2pt,top=2.5pt,bottom=2pt]
\textbf{RQ\textsubscript{2}}: \textit{
SiL exhibits significant actuation mismatches in throttle, steering, and braking, with errors exceeding 30–70\%, while ViL drastically reduces these gaps, achieving over 90\% improvement in trajectory alignment and more accurate speed and braking responses.
}
\end{tcolorbox}

\subsection{RQ\textsubscript{3} (perception validity and gap)}

\autoref{tab:rq3} presents the results for perception validity and gap.
Concerning obstacle alignment, at 40 cm, IoU is high (0.81 for one obstacle, 0.75 for two), and LiDAR error remains low (0.03 m and 0.08 m, respectively), indicating accurate mapping near the vehicle. As distance increases, alignment degrades: at 160 cm, IoU for a single obstacle drops to 0.68 (a 16\% decrease), and LiDAR error rises to 0.12 m, nearly 4$\times$ higher. For two obstacles, degradation is more severe: IoU falls to 0.29 (61\% reduction), and LiDAR error reaches 0.14 m (75\% increase). These trends reflect how distance amplifies projection errors and how FoV edge positioning, especially with lateral placement, affects visual alignment, likely due to lens distortion and radial calibration mismatch.

\begin{table}[t]
\caption{RQ\textsubscript{3}: Perception input validity and gap results.}
\label{tab:rq3}
    \centering
    \footnotesize
    \setlength{\tabcolsep}{9.2pt}
    \renewcommand{\arraystretch}{1.1}
        \begin{tabular}{lcccc}
        \toprule
        Obstacles Validity & \multicolumn{2}{c}{Camera IoU} & \multicolumn{2}{c}{Lidar Dist (m)}\\
         & 1 obs. & 2 obs. & 1 obs. & 2 obs.\\
         \midrule
        40cm & 0.81 & 0.75 & \textbf{0.03} & 0.08\\
        80cm & 0.73 & 0.59 & 0.03 & 0.05\\
        120cm & \textbf{0.90} & 0.42 & 0.06 & 0.04\\
        160cm & 0.68 & 0.29 & 0.12 & 0.14\\
        \midrule
        Lane Validity & \multicolumn{4}{c}{Lanes Distance (pixels)}\\
        & \multicolumn{2}{c}{CW}& \multicolumn{2}{c}{CCW}\\
        \midrule
        center lane & \multicolumn{2}{c}{9.71 ± 6.62} & \multicolumn{2}{c}{5.40 ± 3.24} \\
        left lane & \multicolumn{2}{c}{\textbf{5.08 ± 5.81}} & \multicolumn{2}{c}{\textbf{1.03 ± 1.32}}\\
        left margin & \multicolumn{2}{c}{6.05 ± 7.09} & \multicolumn{2}{c}{5.32 ± 4.02}\\
        right lane & \multicolumn{2}{c}{17.16 ± 27.18} & \multicolumn{2}{c}{6.46 ± 3.71} \\
        right margin & \multicolumn{2}{c}{12.09 ± 7.78} & \multicolumn{2}{c}{12.97 ± 4.18}\\
        \midrule
        Perception Gap & \multicolumn{2}{c}{SiL$\Delta$} & \multicolumn{2}{c}{MR$\Delta$}\\
        \midrule
        \textbf{Camera} & Obs. & Lanes & Obs. & Lanes\\
        Corr. Coeff. (↑)        & -0.38 & 0.26 & \textbf{0.91} & \textbf{0.76}\\
        Hist. Int. (↑ e-02)    & 0.34 & 0.40 & \textbf{7.50} & \textbf{6.90}\\         
        LBP(↑ e+04)   & 4.28 & 4.49  & \textbf{7.55} & \textbf{7.30}\\
        PSNR (↑ e-01)                           & 2.78  & 2.81  & \textbf{3.99} & \textbf{3.57}\\
        SSIM (↑)                           & 0.21  & 0.23  & \textbf{0.93} & \textbf{0.81}\\
        NMI (↑)                            & 0.86  & 0.81  & \textbf{0.94} & \textbf{0.89}\\
        IFD (↓)                            & 1.74 & 1.70 & \textbf{1.69} & \textbf{1.58} \\ 
        KL Div. (↓)                  & 2.62  & 2.12  & \textbf{0.01} & \textbf{0.48}\\
        MSE (↓)                            & 106.20  & 98.76  & \textbf{9.12} & \textbf{21.94}\\
        Perceptual D. (↓)            & 17.39  & 19.57  & \textbf{4.56} & \textbf{8.83}\\          
        GLCM (↓)       & 465.65  & 533.19 & \textbf{9.12} & \textbf{181.80}\\
        WD (↓ e-3)       & 1.34  & 2.09 & \textbf{0.18} & \textbf{0.31}\\
        \midrule
        \textbf{Lidar} &\multicolumn{2}{c}{Obstacles}&\multicolumn{2}{c}{Obstacles}\\
        Max Distance$\epsilon$ (m ↓) & \multicolumn{2}{c}{1.06 ± 0.39} & \multicolumn{2}{c}{\textbf{0.71 ± 0.43}}\\
        Mean Distance$\epsilon$ (m ↓) & \multicolumn{2}{c}{0.10 ± 0.03} & \multicolumn{2}{c}{\textbf{0.03 ± 0.03}}\\
        Std Distance$\epsilon$ (m ↓) & \multicolumn{2}{c}{0.17 ± 0.06} & \multicolumn{2}{c}{\textbf{0.09 ± 0.08}}\\
        \bottomrule
        
        \end{tabular}
\end{table} 

Interestingly, IoU is not strictly monotonic with distance: for single obstacles, it peaks at 120 cm (0.90), suggesting optimal perception when objects are centered and fully within the frame. In contrast, lateral placement consistently worsens results, even when close (e.g., 0.75 at 40 cm vs. 0.59 at 80 cm, 0.42 at 120 cm). These effects are mostly observed at peripheral placements or near-range extremes. Within the typical operating range (40–120 cm with centered FoV), IoU remains above 0.73 and the LiDAR error is below 6 cm, confirming perception accuracy under correct driving conditions.

Concerning lane alignment, lane perception is most accurate in the left lane during counter-clockwise (CCW) runs (1.03 pixels) and least accurate in the right lane during clockwise (CW) runs (17.16 pixels). Center and left-margin lanes show consistent performance across both directions, while right-side lanes degrade more sharply, likely due to imperfect camera placement representation. For example, center-lane error remains moderate at 5.40 pixels (CCW), while right-margin error increases to 12.09 pixels (CW) and 12.97 pixels (CCW). These patterns show that alignment is best near the image center and deteriorates significantly with lateral offset. Even so, within functional bounds, centered and mid-FoV-pixel-level alignment remains within usable error margins, supporting reliable lane detection under typical operation.

Concerning camera perception, SiL shows poor alignment with real-world input. SSIM is low, 0.21 for obstacles and 0.23 for lanes, indicating a mismatch with the E2E model's training distribution and explaining its degraded behavior in nominal scenarios (\textbf{RQ\textsubscript{1}}).
MR significantly improves visual fidelity: SSIM increases to 0.93 (obstacles) and 0.81 (lanes), with correlations of 0.91 and 0.76. Perceptual distance drops by 74\% (obstacles) and 55\% (lanes), LBP similarity nearly doubles, and KL divergence falls from 2.62 to 0.01 (obstacles) and 2.12 to 0.48 (lanes). Across all evaluated metrics, MR consistently reduces the perception gap and aligns more closely with real-world input, reflecting the observed behavioral improvements.

Concerning LiDAR perception, MR also outperforms SiL. 
For obstacle detection, SiL shows a mean depth error of 0.108~m, a maximum of 1.068~m, and a standard deviation of 0.179~m. 
MR reduces these to 0.039~m (–64\%), 0.713~m (–33\%), and 0.090~m, respectively, yielding higher accuracy.  

Overall, these results confirm that SiL introduces substantial perception gaps, particularly for camera input, which contribute to behavior failures. 
MR narrows these gaps, better aligning sensor inputs with real-world distributions and enabling more accurate behavior replication.

\begin{tcolorbox}[boxrule=0pt,frame hidden,sharp corners,enhanced,borderline north={1pt}{0pt}{black},borderline south={1pt}{0pt}{black},boxsep=2pt,left=2pt,right=2pt,top=2.5pt,bottom=2pt]
\textbf{RQ\textsubscript{3}}: \textit{
MR significantly reduces perceptual discrepancies compared to SiL, improving camera SSIM from 0.21 to 0.93 and reducing LiDAR error by over 60\%, resulting in realistic sensor inputs that closely match real-world data and enable accurate system behavior replication.
}
\end{tcolorbox}

\subsection{Threats to Validity}\label{sec:ttv}

\subsubsection{Internal validity}

We compared all ADS under identical parameter settings.
One threat to internal validity concerns our custom implementation. However, this was unavoidable as no similar evaluation frameworks are available, to the best of our knowledge. 
Another threat may be due to our data collection phase and training of ADS, which may exhibit a large number of misbehaviors if trained inadequately or with poor-quality data. 
\changed{We mitigated this threat by training and fine-tuning the best publicly available driving models, which performed consistently in nominal RW conditions. 
The RW-trained E2E model completed scenarios N1 and N2 in all five RW runs without failure (Table~1, Nominal–E2E–Real), indicating adequate training. 
The LiDAR-based modular pipeline failed in N2 due to late-appearing obstacles, as its perception relies on deterministic clustering rather than learned models; however, the ablation study shows the planning module operates correctly when decoupled from LiDAR inputs. 
Failures in scenario G stem from domain shift rather than poor training: the E2E ADS succeeds in SiL but fails in RW, which would not occur if the model were fundamentally incapable of solving the task. 
Overall, these results confirm that our comparisons are not confounded by under-trained or low-quality models.
}

\subsubsection{External validity}

We used a limited number of ADS architecture models in our evaluation, which we mitigated by covering representative ADS architectures. We considered only two physical tracks and a scaled-down platform, which may not capture all real-world physics (e.g., suspension dynamics or vehicle mass distribution). However, our goal is not to model absolute performance but to assess relative fidelity across modalities. Donkey Car was used as a proxy for full-size ADS also in previous studies~\cite{2021-01-0248,9412011,viitala,DBLP:journals/corr/abs-1909-03467,2023-Stocco-TSE,2023-Stocco-EMSE} and uniquely satisfies our requirements for studying transferability between simulated and real-world testing of ADS. Other platforms, such as DeepPiCar~\cite{bechtel2018deeppicar} and JetRacer~\cite{jetracer}, lack integrated simulators; Roboracer~\cite{f1tenth} offers only low-fidelity physical simulations that do not realistically capture real-world driving dynamics. AWS DeepRacer~\cite{aws-deepracer} is tightly integrated with AWS infrastructure and is primarily designed for reinforcement learning use cases, which are outside the scope of this work. In contrast, Donkey Car has been successfully adopted in numerous real-vehicle autonomous driving studies~\cite{2021-01-0248,9412011,viitala,DBLP:journals/corr/abs-1909-03467,2023-Stocco-TSE,2023-Stocco-EMSE,li2024autonomousdrivingsmallscalecars,CALEFFI2024271,Mokhtarian}, making it a practical and cost-effective experimentation platform.

We acknowledge that the availability of multiple and diverse tracks and obstacle configurations would be desirable. However, our selection of scenarios meant to isolate and evaluate core ADS capabilities, namely lane-keeping and obstacle avoidance, under controlled, repeatable conditions. Hence, generalizability to other physical settings or RC platforms might not hold or may hold partially. \changed{We use a scaled-down platform, which may not capture all real-world physics (e.g., suspension dynamics or vehicle mass distribution). However, our goal is not to model absolute performance but to assess relative fidelity across modalities.}

\subsubsection{Reproducibility} 

All software artifacts and results are available in our replication package and appendix~\cite{replication-package}. To replicate our study, however, two physical assets are needed, i.e., a Donkey Car and a tracking system.
\section{Discussion}

\changed{
\subsection{Dissecting and Addressing the ADS Reality Gap}
Our study highlights persistent differences between SiL, ViL, MR, and RW testing, underscoring the need to improve transferability across these environments. While simulation remains indispensable for ADS validation, our results show it is insufficient in isolation. Two major challenges were observed, namely the \textit{perception gap} and the \textit{actuation gap}.}

\changed{
To mitigate the former in SiL, without relying on physical vehicles, recent neural rendering approaches have been proposed. Examples include generative AI for translating simulated into photorealistic images~\cite{amini-gap,2023-Stocco-TSE,2023-Stocco-EMSE,biagiola2023better,lambertenghi_ICST24,10880098} and diffusion models for generating realistic operational design domains~\cite{2025-Baresi-ICSE,baresi2025dillemadiffusionlargelanguage,2025-Guo-arxiv}. Some are already integrated in simulators such as NVIDIA Omniverse. While promising, these methods still suffer from correctness issues (e.g., artifacts, hallucinations), increase runtime cost, and leave the actuation gap and simulator unreliability~\cite{AfzalSimulation21,AminiFlaky2024} unaddressed.  
}

\changed{
On the other hand, the actuation gap is better addressed through hybrid testing. ViL and MR, as evaluated in this work, virtually eliminate actuation mismatches, with MR additionally reducing perception errors. However, these setups are not a substitute for SiL experiments as running tests on real vehicles and hardware remains costly and resource-intensive, even if less than RW. Instead, we view ViL and MR as complementary, highlighting the need to develop strategies to prioritize which scenarios merit RW execution.
}

\changed{
Our results highlight meaningful differences across ADS types. In SiL, E2E systems appear under-confident while modular pipelines appear overconfident. Under generalization scenarios that introduce distribution shifts, the trend reverses, with E2E failing in RW while modular ADS proves more robust. 
To address this, we suggest a staged strategy: SiL for early validation and coverage, ViL for refining control, and MR for perception fidelity. At the same time, MR setups are more expensive and harder to deploy than SiL, motivating our release of a modular and well-engineered framework. 
}

\changed{
Finally, our findings must be interpreted with respect to the experimental setup: the vehicle is small-scale, tracks and scenarios are simple, and our open-source simulator, particularly its rendering, is not fully photorealistic. These factors may have amplified the transfer gap, especially for perception-heavy ADS pipelines. Nonetheless, we expect similar issues, though at different magnitudes, even with industrial-grade simulators and full-scale vehicles. 
}

\changed{
\subsection{Implications for testing, debugging, and monitoring}
Hybrid testing offers promising opportunities to improve validation, debugging, and monitoring of ADS. 
For \textit{validation}, we propose the use of ViL and MR to replay critical or failure-inducing cases originally detected in SiL~\cite{Gambi:2019:ATS:3293882.3330566,wu2025multiobjectivereinforcementlearningcritical}. This approach supports more trustworthy assessments of safety and behavior and enables more representative test generation~\cite{2025-Sorokin-arxiv}.  
}
\changed{
From a \textit{debugging} perspective, MR provides a valuable trade-off. Unlike pure simulation, it avoids artificial failures; unlike RW testing, it enables repeatable, cost-effective, and versatile experimentation with physical vehicles. This makes MR particularly suitable for analyzing rare or complex issues such as braking delays or occluded pedestrian responses, although the transferability of such insights to full-scale RW vehicles remains open.  
}
\changed{
For \textit{monitoring}, ViL and MR allow evaluation under realistic latencies, noise, and actuation constraints. Since many state-of-the-art monitoring tools~\cite{2022-Stocco-ASE,2020-Stocco-ICSE,2024-Grewal-ICST,deepsafe} are assessed only in simulators, hybrid environments help bridge this gap by enabling controlled crash reproduction, evaluation under adverse conditions, and the training of more robust monitoring systems. Early steps in this direction include the works of Ayerdi et al.~\cite{marmot} and Huang et al.~\cite{huang2025reactruntimeenabledactivecollisionavoidance}.
}
\section{Related Work}\label{sec:related}

\subsection{Reality Gap Assessment Studies}\label{sec:reality-gap-assessment}

The reality gap has been subjected to active research in many fields, including robotics, automotive, and artificial intelligence. For a comprehensive survey, we refer the reader to Hu et al.~\cite{gap-survey}, and contributions in the software engineering community~\cite{10.1007/s10515-025-00523-7,survey-lei-ma, AfzalSimulation21, 2023-Stocco-TSE, fse-survey-robotics, icst-survey-robotics}.
Concerning empirical studies on the reality gap, Stocco et al.~\cite{2023-Stocco-TSE} compare ADS lane-keeping models in simulated and physical environments, highlighting critical shortcomings that contribute to the gap. In this work, beyond assessing the gap between SiL and RW, we also evaluate mitigation strategies such as ViL and MR. Similarly, Gao et al.\cite{10.1145/3597503.3639191} propose MultiTest, a physically-aware object insertion framework for testing the robustness of fusion-based perception systems, while Gao et al.~\cite{10.1145/3611643.3616278} outline key challenges in benchmarking AI-enabled multi-sensor fusion across diverse conditions. However, prior studies focus mainly on perception robustness in simulation and stop short of full-system evaluation. To our knowledge, our work offers the first unified, system-level assessment of ViL and MR as reality-gap mitigation strategies, analyzing their impact on perception, actuation, and behavior during live execution in SiL, MR, and RW under consistent, controlled conditions.
However, the transferability to real vehicles is not assessed.

\subsection{Reality Gap Mitigation Studies}\label{sec:reality-gap-mitigation}

Concerning solutions to mitigate the gap, researchers have proposed a variety of strategies. One common approach involves the use of digital twins, which aim to replicate real-world vehicle dynamics and sensor characteristics with high fidelity~\cite{DTwins}. Alternatively, search-based tuning of simulator parameters can be employed using real-world logs~\cite{khatiri2023simulation}.
Another technique is domain randomization, which improves generalization by varying environmental parameters such as lighting, weather, or road conditions during training~\cite{domain_rand,lambertenghi_ICST25}.
However, domain randomization and adversarial training are typically applied to models trained only in simulation. This is not the case for ADS, as they are trained on real-world data. 

To address the perception gap, researchers have explored the use of generative AI to translate simulated data into photorealistic images~\cite{amini-gap,2023-Stocco-TSE,2023-Stocco-EMSE,lambertenghi_ICST24,10880098}, LiDAR point clouds~\cite{lidar_gap}, and methods for synthesizing operational design domains with high visual fidelity~\cite{2025-Baresi-ICSE,baresi2025dillemadiffusionlargelanguage,Lindstrom_2024_CVPR,Tonderski_2024_CVPR}. While existing methods enhance perceptual realism, they typically run offline at the model level or in SiL, targeting single sensor modalities. In contrast, our work evaluates perception fidelity during live, system-level execution across SiL, MR, and RW, directly comparing how perceptual gaps affect the ADS behavior.

\section{Conclusions}\label{sec:conclusions}

This paper presents a comprehensive empirical study of the reality gap in autonomous driving, analyzing Software-in-the-Loop (SiL), Vehicle-in-the-Loop (ViL), Mixed-Reality (MR), and real-world execution across both modular and end-to-end driving systems. Our goal is to isolate the dimensions of the reality gap, namely, behavior, actuation, and perception, and assess how each modality reflects real-world behavior.

Our findings reveal that the reality gap is multifaceted and modality-dependent. SiL often misrepresents system behavior, failing when the real system succeeds, and vice-versa. ViL mitigates actuation errors but leaves perception gaps unresolved. MR, with simulated obstacles and RW perception, is the only testing approach that more consistently captures both perceptual and behavioral fidelity, matching real-world outcomes in both nominal and generalization scenarios. 

Our openly available framework and results offer a foundation for the development of next-generation ADS testing and validation solutions, promoting the adoption of hybrid setups that combine SiL, ViL, and MR to enable efficient, scalable ADS evaluation.

\section*{Acknowledgements}
\addcontentsline{toc}{section}{Acknowledgements}
This research was funded by the Bavarian Ministry of Economic Affairs, Regional Development and Energy.

\balance
\bibliographystyle{IEEEtran}
\bibliography{paper}

\end{document}